\documentclass[conference]{IEEEtran}

\usepackage[backend=bibtex,style=verbose-trad2,style=numeric,
citestyle=numeric]{biblatex}
\IEEEoverridecommandlockouts
\usepackage{amsmath,amssymb,amsfonts}
\usepackage{algorithmic}
\usepackage{graphicx}
\usepackage{textcomp}
\usepackage{xcolor}
\usepackage{listings}
\def\BibTeX{{\rm B\kern-.05em{\sc i\kern-.025em b}\kern-.08em
    T\kern-.1667em\lower.7ex\hbox{E}\kern-.125emX}}
\usepackage{sansmath}

\bibliography{paper}

\definecolor{codegreen}{rgb}{0,0.6,0}
\definecolor{codegray}{rgb}{0.5,0.5,0.5}
\definecolor{codepurple}{rgb}{0.58,0,0.82}
\definecolor{backcolour}{rgb}{0.95,0.95,0.92}

\lstdefinestyle{mystyle}{
    backgroundcolor=\color{backcolour},   
    commentstyle=\color{codegreen},
    keywordstyle=\color{magenta},
    numberstyle=\tiny\color{codegray},
    stringstyle=\color{codepurple},
    basicstyle=\ttfamily\footnotesize,
    breakatwhitespace=false,         
    breaklines=true,                 
    captionpos=b,                    
    keepspaces=true,                 
    numbers=left,                    
    numbersep=5pt,                  
    showspaces=false,                
    showstringspaces=false,
    showtabs=false,                  
    tabsize=2
}

\begin{document}

\title{Towards Auditable Distributed Systems\\
\thanks{*This work was supported by the Federal Ministry for Economic Affairs and Climate Action in the RTAPHM project, grant 20X1736M (Lufo V-3). We thank also Dr. Ulrich Schoepp for support in the implementation, valuable feedback and discussions on the topic. }
}

\author{\IEEEauthorblockN{ Lev Sorokin*}

\IEEEauthorblockA{
fortiss, Research Institute of the Free State of Bavaria, Guerickestraße 25, 80805 Munich, Germany \\
sorokin@fortiss.org}
}
\maketitle

\begin{abstract} The emerging trend towards distributed (cloud) systems (DS) has widely arrived whether in the automotive, public or the financial sector, but the execution of services of heterogeneous service providers is exposed to several risks. Beside  hardware/software faults or cyber attacks that can influence the correctness of the system, fraud is also an issue. In such case it is not only important to verify the correctness of the system, but also have evidence which component and participant behaves faulty. This makes it possible, e.g. to claim for compensation after systems' execution but also to assure information for verification can be trusted. The main goal of our research is to assure the monitoring of DS based on auditable information. We follow a decentralized monitoring strategy and envision a distributed monitoring approach of system properties based on \textbf{distributed logic programs} that consider auditability. The expected contribution of this work is to establish with the application of our framework the mutual trust of distributed parties, as well as trust of clients in the systems execution. 
 We showcase our ideas on a DS for booking services with unmanned air vehicles.

\end{abstract}


\section{Introduction}

Distributed systems (DS) have become omnipresent and find applications in, e.g. the automotive domain considering Vehicle-To-Vehicle communication \cite{V2V}, as distributed payment systems \cite{uber} or for the collaborative training of machine learning models in the e-government sector \cite{afml}. DS are prone to hardware or software errors, network failures, as well as cyber-attacks, that change the configuration of a system, its database and modify, delay or produce additional messages sent between system components \cite{faults}. Especially, when components are distributed over different organizational domains and maintained by distinct providers, it is difficult to assure that a system meets its specification when executed systems code is not accessible and fraudulent parties might have an incentive to cheat.

One approach to verify that a DS execution is conformant to its specification, is to monitor the system during runtime. Runtime verification (RV) \cite{rv_intro} is applied in general, when the system models are not given or the states are too complex (state explosion) to perform exhaustive testing. In RV a monitor receives a property specification (mostly in temporal logics) and verifies the validity of the property given a set of system actions provided at runtime (usually an event trace). On the one hand, it is important to assure the correct execution of such a system, on the other hand one needs to have convincing evidence which component is faulty.  
For instance a provider of a flight booking service might decline having received a payment by a travel agency. Here it is necessary to have evidence in hindsight to claim compensation for incorrect behaviour, but also at the time of system processing to mitigate resulting damage (e.g. to the customer). This is especially important in safety and security-critical applications. In this regard, \textit{accountability} of parties is required, which is described
in \cite{auditability09} as a property of a system where an internal or external party is supposed ``[...] to blame protocol participants in case of \textit{misbehaviour}".  A misbehaviour can be considered  when a desired goal in the system is not met, e.g. as in the previous example when a party claims having received information that is different from sent information. 
In this paper we are focusing on \textit{auditability}, what in contrast does not require the blaming part, it is defined as following \cite{auditability09}: a system is ”[...] auditable, if at any audit point, an impartial judge is satisfied with the evidence produced by the [system]”. We consider a DS is auditable, if every participant and also external observer are able to verify whether a property of the system is valid or not.

While many works do exist for the RV of DS, as e.g. \cite{rvMicroservicesDesignToRuntime, cotroneo2020runtime}, only a few approaches are addressing auditability of processed events. Auditability is required also for monitoring to assure that the reasoned verdicts are correct. The primary building blocks for an auditable system incorporate an append-only tamper-proof ledger, to avoid repudiation of committed actions and a verification mechanism to detect changes of logged data \cite{accountableCloud}. The combination of monitoring in large-scale DS, time-consuming auditing mechanisms and logging transactions  can create a considerable overhead. For now papers that consider auditability rely on distributed ledgers and regard mostly not scalable applications such as supply chain workflows as well have a complex deployment \cite{UntrustedBusinessProcessBlockchain, Prybila2020RuntimeVF}. We consider that a framework is required to deal with: (1) A distributed system consisting of black box components, (2) Enforcement of security (e.g. authorization) and workflow properties, (3) Identification of the responsible parties when a property is violated and a system execution that can be reviewed later by an external party, and (4) A large number of transactions in the DS. In this paper we intend to discuss an approach towards a framework that captures the aforementioned demands.

\section{Use Case}
\begin{figure}
    \centering
    \includegraphics[scale=0.27]{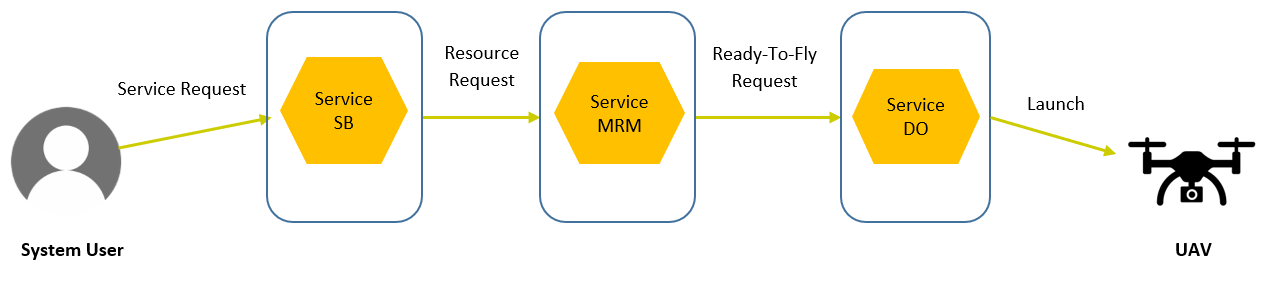}
    \caption{Distributed booking system for UAVs}
    \label{fig:rtaphm}
\end{figure}
We motivate our approach with a use case from the RTAPHM project \cite{rtaphm}.
Unmanned air vehicles (UAVs) have gained high popularity in the last few years, their applications range from cargo transportation, to search-and-rescue operations or environmental monitoring \cite{UAVApplications}.
We consider as an example a distributed platform for scheduling UAVs for the delivery of a specific type of cargo, namely organs \cite{droneOrganTransport}. It consists of heterogeneous web service components (s. Figure \ref{fig:rtaphm}), hosted by different providers: the Service Broker (SB) for checking the correctness and authorization of requests, the Multi-Resource-Manager (MRM) for managing available resources, as well as the Drone Operator (DO) for executing services. Further it consists of personnel responsible for the preparation of UAVs for the flights, different kind of drones and trivially users that send service requests and confirm booking options. In a fault free scenario a user requests an organ transport from destination A to destination B; the SB checks the request and forwards it to the MRM to determine an available UAV, as well as assets as transport boxes, and personnel to meet the request constraints (e.g. as delivery-time, organ size).
The user receives as a response a set of booking options together with prices. After he has selected one option the request turns into a mission, the personnel is triggered to prepare the drone. If the preparation is completed, the DO receives a ready-to-fly (RTF) request to launch the drone. 

We assume a specification is given, in terms of which APIs the system components have and which communication workflows in the distributed platform are allowed. 
In presence of a faulty behaviour, we consider, e.g. the following cases that can occur in the system:

\begin{enumerate}
\item \textit{
An attacker is passing artificial RTFs to DO to launch UAVs. The DO schedules a UAV without a request by the user, because the DO's security mechanism is faulty.
} 
\item \textit{
An RTF-message is sent delayed to the DO due to a hardware fault at MRM, as a consequence the organ is damaged. } 
\end{enumerate}

If the presented DS is auditable and related properties are monitored it is possible to detect that no party has received any booking and that therefore the RTF message is malicious. This makes it possible to blame DO for its malfunction and at the same time reason about the system with verifiable information.


\section{State-of-the-Art}

This section gives a brief summary of related work covering the runtime monitoring of distributed systems in connection of auditable execution. 

In the business process management domain \cite{Prybila2020RuntimeVF} has applied blockchain for a choreography based RV of business processes to penalize system participants when they behave faulty or justify a payment for work done to achieve a business goal. Their approach cannot enforce policies and is limited to business processes and applications on special domains like logistics or banking because it is not scalable.

Another approach \cite{peerreview} is focusing on the detection of Byzantine faults based on an auditing protocol using local tamper-proof logs. The idea is that all components do locally log all sent and received messages and the application events, replay the execution of other components and compare the logs to detect deviations. Their approach assumes that components behaviour is determinstic and it does not require a specification given, but requires shared component code across the system, what is problematic in a DS with parties of distinct domains.  The authors of \cite{accountableCloud} discuss requirements and present building blocks to make cloud services auditable, but the performance of such a framework that uses this mechanisms is unclear and auditing is considered to be done after systems execution.

Our work is based on the concept presented in \cite{afml} for accountable collaborative training of machine learning models, but the difference is that auditing is performed after training of the model and not at runtime in contrast to what is envisioned in our approach.

Some other work is limited to monitoring business process using Blockchain \cite{UntrustedBusinessProcessBlockchain}. Their approach is not scalable, why it supports only use cases related to supply chains. 


\section{Research Questions}

Studying the existing work and considering our requirements this leads us to the main research question we want to answer: \textbf{How can we make distributed systems auditable?}
\\

On a finer level we state the following research questions:

\begin{itemize}

    \item \textbf{RQ1. Given a DS and its specification, how can we automatically and application-independently make DS auditable?}
     The approach to assure that a DS execution is auditable should be application-independent and not require parties to implement their own monitoring solution, but rather to have a specification from which the monitoring system that provides the auditability is automatically derived.
    
    \item \textbf{RQ2. What is the performance impact (i.e. request delay, throughput) when we perform monitoring of system properties based on auditable information during runtime synchronously?}   It is unclear which performance impact it yields when applying monitoring with auditable events at runtime, and not as usual, after system execution.
    We consider to research this for the case the system execution is halted for the auditing process. 
    
    \item \textbf{RQ3. How resource intensive (CPU, storage) is the monitoring of system properties based on auditable information during runtime?} Similar as stated in RQ2, a runtime monitoring that considers auditability requires processing capabilities that might have an impact on the deployment of the monitoring solution.
\end{itemize}    



\section{Approach}
\label{sec:approach}

Our approach has a connection to the Open Policy Agent (OPA) \cite{opa}, an authorization framework, that allows to specify policies in a declarative way that are decoupled from code and enforce those in a DS, primary cloud native environments. We want to extend OPA to take auditability of processed events at runtime as well as after the systems execution into account. To achieve this we consider to use a distributed monitoring architecture as shown in Figure \ref{fig:evid_architecture}, where 1) a monitor is attached to each system component, which logs relevant components information (e.g. ingoing/outgoing messages) in a scalable common log tamper-proof and undeniable, and 2) it can exchange events with other components and monitor properties based on this recorded information. A third party can access the recorded information in the common log to reason about the systems state. A monitor can operate in two different ways: it can verify a property without blocking the execution (``Trust, then Verify"-manner), or it verifies first that a system is in a correct state wrt. the observed event (``Verify, then trust"). 

Further, each monitor receives a set of properties to be monitored. We consider to have such a decentralized monitor setup to avoid a single point of failure (SPOF) and to reduce the data flow between a component and the log, since not all observed information is to be sent to a central monitoring unit but rather only verdicts. This is similar to choreography-based RV \cite{Francalanza2018}. 

To enable the monitoring of properties considering the presented architecture we need to execute following steps:

\begin{enumerate}

    \item Identification of a language to specify the systems events and properties that incorporate principals identities

    \item Identification and evaluation of different approaches for auditing events at runtime
    
    \item Examine an approach to automatically generate monitors for auditing events based on the results from previous steps
        
    \item Identification and evaluation of integrating secure logging solutions considering scaliabality in our approach
\begin{figure}[h!]
    \centering
    \includegraphics[scale=0.47]{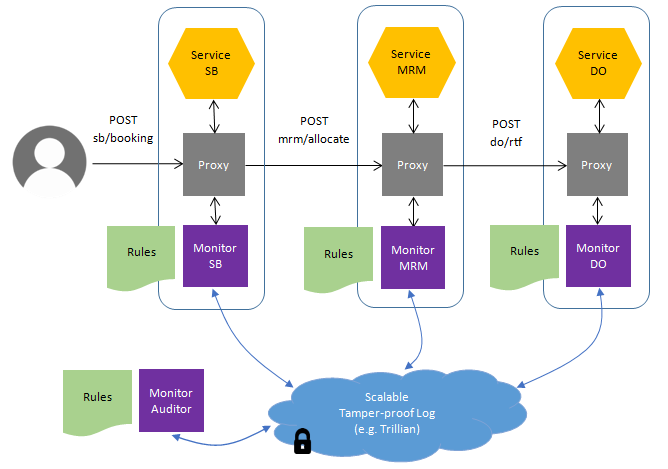}
    \caption{Monitoring architecture based on auditable events for the distributed UAV booking use case}
    \label{fig:evid_architecture}
\end{figure}
\end{enumerate}
We describe the steps in the following:

(*Step 1) In the first step a specification language is required to specify the to monitorable properties. Using this language one should be able to incorporate principals identity information, to express individual claims. We have initially selected a language based on the declarative, rule-based logic programming language Datalog \cite{datalogCeri89}. 
Our specification is based on the idea of the attestation logic Cyberlogic  \cite{cyberlogic} that considers principal information in a claim and assumes to have a PKI infrastructure given. 
We have selected a datalog based specification language due to the following reasons: 1) it has been proven to be suitable for the specification of invariants in distributed systems \cite{DeclarativeCloud}, 2) rule based languages are considered as intuitive for specifying policies as shown in \cite{DeclarativeCloud} or by Rego as used in OPA. Using this language we can model the components/principals' knowledge about its state in the way ``\texttt{principal} \texttt{attests} \texttt{claim}". For instance, consider Listing 1 that specifies a set of rules that are based on such facts to detect an incorrectly submitted RTF (property 2). The first and the second rule indicate that booking and RTF information is derived from a request.

    \begin{lstlisting}[language=Prolog, caption=Specification of property 2 with datalog-based rules for the detection of incorrectly submitted RTF requests]
'DO' attests ready_to_fly(Id,T,C) :- 
'DO' attests postRequest('\ready_to_fly',Id,T,C).
  
'SB' attests booking(Id,T,C)  :-   
'SB' attests postRequest('\booking_request',Id,T,C).
    
exists_booking_before(Id,End,C) :- 
    'SB' attests booking(Id,T,C), T < End.
    
forbidden(Id):- 
    not exists_booking_before(Id,T,C1),
	'DO' attests ready_to_fly(Id,T,C2).
\end{lstlisting}
    
The parameter Id relates to a specific booking session, while T is the timestamp and C the data provided in the request. 

(*Step 2) Having a specification of a property to be monitored, one needs to develop a monitoring solution that assures the integrity of the provided data at runtime. We consider two necessary mechanisms for this:

\textbf{Signing:} All parties are part of a public key infrastructure \cite{pkiKohnfelder1978TowardsAP}. Similar to the approach in \cite{AuditingWorkflowsDec}, a sent message is signed with the principals private key, and decoded by using its public key, to know the origin of the message.

\textbf{Proof search:} When a party1 sends a message to party2, beside using a signed message with the sender's private key, party2 checks for the justification of the event. For instance, if MRM sends an RTF to the DO, it has to be checked whether there is some justification for the RTF message in terms of a booking confirmation and a service request. The evidence can be provided by a common log, or by the sender. The performance is to be investigated of both approaches. 



(*Step 3) Given a formal specification of the properties that should be verified in the language of step1, the goal is automatically to generate monitors to enable auditing (step2) to avoid writing manually a monitor for each component. Given a datalog specification, we consider to have as the input specification rules for the whole system, and as the output a subset of such rules for each monitor.

(*Step 4) A logging mechanism is required to enable that all system events are tamper-proof and undeniable, so that they can also be reviewed after execution. Secure storage solutions based on distributed ledgers (DL) as hyperledger \cite{hyperledger} exist, but also centralized as Trillian \cite{trillian}, which is more scalable and requires less resources than DL. Nevertheless centralized ledgers pose a SPOF. Auditing of the secure log itself is then required to avoid that the log presents a record to different principals differently. The performance is to be evaluated, since stored information at each log has to be compared, when, e.g. several centralized logs are deployed in parallel.

\section{Current State}

We have implemented a prototype to have a starting point to investigate our research questions and we are currently working on step 2. For this, we have implemented a monitor with a Datalog engine to execute rules, that represent system properties to reason about the system state. One issue we have encountered using Datalog is to handle the large amount of facts in the local storage of a monitor, since
deleting facts can cause an issue when a required fact to apply a rule is not available. This can lead to a resolution that a systems property is considered incorrectly as valid. Therefore, we are working on a strategy based on revisions where facts in the local database are only stored of the latest revision as proposed by \cite{Dedalus}. As a secure log we have first used hyperledger and finally Trillian. The results have shown that as expected hyperledger limits the maximal number of transactions to 1000/s, what does not fit our requirements. The distributed services from the use case have been provided as Docker images together with an API description and self-signed certificates for testing. We have deployed the docker containers in an one node kubernetes cluster and specified the rules for the monitors, derived from the API description and the business logic given. 
Finally we deployed a monitor per service to intercept requests/responses using a service mesh and restrict access to endpoints depending on the systems state. 


\section{Discussion and conclusion}

We have presented a concept to make distributed systems auditable not only after execution but also at runtime by using distributed monitors that record and exchange undeniable and tamper-proof events to reason about the systems state. Our approach, based on distributed logic programs, supports not only to verify properties in hindsight, but also to enforce desired behaviour using auditable information. 

The proposed solution has several limitations and challenges. One of the limitations is to deal with a threat, where all private keys for signing messages get stolen and an attacker creates all required evidence for an event (attacker knows the contracts). Technical challenges that are not considered are: (1) Which events should be logged, since logging all events is not a feasible solution? (2) How to determine the order of events, when no global clock is considered? Also organizational challenges have to be considered: (1) Trust is required that a party agrees launching a monitor and communication is logged correctly  and (2) information processed by a component could be confidential, and only accessible by authorized components; this can make it impossible to construct a proof tree for an event. 
Nevertheless, we believe that further successful research and implementation of our ideas will establish trust between parties in DS but also of users in cloud systems and contribute to the security and safety in the distributed systems' execution.



\printbibliography

@inproceedings{rv_intro,
author = {Bartocci, Ezio and Falcone, Yliès and Francalanza, Adrian and Reger, Giles},
year = {2018},
month = {02},
pages = {1-33},
title = {Introduction to Runtime Verification},
booktitle = {Lectures on Runtime Verification}
}

@online{droneOrganTransport,
author = {},
url = {https://www.cbc.ca/news/canada/toronto/first-lung-transplant-drone-1.6208057},
urldate = {2022-01-11},
year = {},
month = {},
pages = {},
title = {Drone Organ Transport},
volume = {},
journal = {}
}

@inproceedings{faults,
  author={P. {Thambidurai} and  {You-keun Park}},
  booktitle={Proceedings 7th Symposium on Reliable Distributed Systems}, 
  title={Interactive consistency with multiple failure modes}, 
  year={1988},
  volume={},
  number={},
  pages={93-100}
 }

@misc{cotroneo2020runtime,
      title={Towards Runtime Verification via Event Stream Processing in Cloud Computing Infrastructures}, 
      author={Domenico Cotroneo and Luigi De Simone and Pietro Liguori and Roberto Natella and Angela Scibelli},
      year={2020},
      primaryClass={cs.SE}
}

@inproceedings{peerreview,
author = {Andreas Haeberlen et al.},
title = {PeerReview: Practical Accountability for Distributed Systems},
year = {2007},
publisher = {ACM},
booktitle = {Proceedings of 21st ACM SIGOPS Symposium on Operating Systems Principles},
pages = {175–188},
numpages = {14},
}

@article{Prybila2020RuntimeVF,
  title={Runtime Verification for Business Processes Utilizing the Bitcoin Blockchain},
  author={Christoph Prybila and S. Schulte and C. Hochreiner and I. Weber},
  journal={Future Gener. Comput. Syst.},
  year={2020},
  volume={107},
  pages={816-831}
}

@Inbook{Francalanza2018,
author="Francalanza at al.",
title="Runtime Verification for Decentralised and Distributed Systems",
bookTitle="Lectures on Runtime Verification: Introductory and Advanced Topics",
year="2018",
pages="176--210"
}

@online{accountability,
  author = {},
  title = {},
  year = 2021,
  url = {http://dig.csail.mit.edu/2014/AccountableSystems2014/},
  urldate = {2021-03-10}
}

@inbook{rvMicroservicesDesignToRuntime,
author = {Matteo Camilli et al.},
year = {2018},
month = {02},
pages = {168-173},
title = {Design-Time to Run-Time Verification of Microservices Based Applications}
}

@online{V2V,
  author = {nhtsa.gov},
  title = {V2V-communication},
  url = {www.nhtsa.gov/technology-innovation/vehicle-vehicle-communication},
  urldate = {2022-01-11}
}

@online{uber,
  author = {},
  title = {Ubers Payments Platform},
  url = {https://eng.uber.com/payments-platform/},
  urldate = {2022-01-11}
}

@online{opa,
  author = {},
  title = {OPA},
  url = {https://www.openpolicyagent.org/
  },
  urldate = {2021-06-21}
}

@article{cyberlogic,
author = {Ruess, Harald and Shankar, Natarajan},
year = {2003},
month = {01},
pages = {},
title = {Introducing Cyberlogic},
journal = {National Security Agency’s third High Conﬁdence Software and Systems
Conference}
}

@article{accountableCloud,
author = {Haeberlen, Andreas},
year = {2010},
month = {04},
pages = {52-57},
title = {A case for the accountable cloud},
volume = {44},
journal = {Operating Systems Review}
}

@online{trillian,
title = {Trillian},
url = {https://transparency.dev/#trillian
},
urldate = {2021-12-28}
}

@article{datalogCeri89,
   author = {Stefano Ceri et al.},
   issue = {1},
   journal = {IEEE Transactions on Knowledge and Data Engineering},
   pages = {146-166},
   title = {What You Always Wanted to Know About Datalog (And Never Dared to Ask)},
   volume = {1},
   year = {1989},
}

@online{UAVApplications,
title = {UAV-Applications},
url = {t.ly/-u5r},
urldate = {2021-12-28}
}

@InProceedings{Dedalus,
author="Alvaro, Peter
and Marczak, William R.
and Conway, Neil
and Hellerstein, Joseph M.
and Maier, David
and Sears, Russell",
editor="de Moor, Oege
and Gottlob, Georg
and Furche, Tim
and Sellers, Andrew",
title="Dedalus: Datalog in Time and Space",
booktitle="Datalog Reloaded",
year="2011",
publisher="Springer Berlin Heidelberg",
address="Berlin, Heidelberg",
pages="262--281"
}

@inproceedings{DeclarativeCloud,
author = {Alvaro, Peter and Condie, Tyson and Conway, Neil and Elmeleegy, Khaled and Hellerstein, Joseph and Sears, Russell},
year = {2010},
month = {01},
pages = {223-236},
title = {Boom analytics: exploring data-centric, declarative programming for the cloud}
}

@inproceedings{afml,
author = {Balta, Dian and Sellami, Mahdi and Kuhn, Peter and Sch\"{o}pp, Ulrich and Buchinger, Matthias and Baracaldo, Nathalie and Anwar, Ali and Ludwig, Heiko and Sinn, Mathieu and Purcell, Mark and Altakrouri, Bashar},
title = {Accountable Federated Machine Learning in Government: Engineering and Management Insights},
booktitle = {Electronic Participation: 13th IFIP WG 8.5 International Conference, EPart 2021, Granada, Spain, September 7–9, 2021, Proceedings},
pages = {125–138},
}

@InProceedings{AuditingWorkflowsDec,
author="Nehme, Antonio
and Jesus, Vitor
and Mahbub, Khaled
and Abdallah, Ali",
title="Decentralised and Collaborative Auditing of Workflows",
booktitle="Trust, Privacy and Security in Digital Business",
year="2019",
publisher="Springer International Publishing",
address="Cham",
pages="129--144"
}

@InProceedings{UntrustedBusinessProcessBlockchain,
author="Weber, Ingo
and Xu, Xiwei
and Riveret, R{\'e}gis
and Governatori, Guido
and Ponomarev, Alexander
and Mendling, Jan",
title="Untrusted Business Process Monitoring and Execution Using Blockchain",
booktitle="Business Process Management",
year="2016",
publisher="Springer International Publishing",
address="Cham",
pages="329--347"
}

@online{rtaphm,
title = {RTAPHM-Projekt},
url = {https://www.fortiss.org/forschung/projekte/detail/rtaphm
},
urldate = {2022-06-26}
}

@online{hyperledger,
title = {hyperledger},
url = {hyperledger.org},
urldate = {2022-06-26}
}

@report{pkiKohnfelder1978TowardsAP,
  title={Towards a practical public-key cryptosystem.},
  author={Loren M. Kohnfelder},
  institution={Thesis, Massachusetts Institute of Technology, Cambridge},
  year={1978}
}

@inproceedings{auditability09,
author = {Guts, Nataliya et al},
year = {2009},
month = {09},
pages = {168-183},
title = {Reliable Evidence: Auditability by Typing},
volume = {5789}
}

\end{document}